\documentclass[10pt,journal,letterpaper,compsoc]{IEEEtran}

\usepackage{epstopdf}
\usepackage{graphicx}
\usepackage{stmaryrd}
\usepackage{amsmath,amsthm,amssymb}
\usepackage{multirow,url}
\usepackage{color,cite}
\usepackage{algorithm}
\usepackage{algorithmicx}
\usepackage{algpseudocode}
\usepackage{setspace}
\usepackage{rotating}
\newtheorem{lem}{Lemma}

\newtheorem{thm}{Theorem}
\newtheorem{defn}{Definition}

\ifodd 0
\newcommand{\com}[1]{\textbf{\color{red} (COMMENT: #1)}} %comment of the text
\newcommand{\comg}[1]{\textbf{\color{green} (COMMENT: #1)}}
\newcommand{\response}[1]{\textbf{\color{magenta} (RESPONSE: #1)}} %response to comment
\else

\newcommand{\com}[1]{}
\newcommand{\comg}[1]{}
\newcommand{\response}[1]{}
\fi
\begin{document}

\title{Decentralized Computation Offloading Game For Mobile Cloud Computing}

\author{Xu Chen, \emph{Member, IEEE} \thanks{The author is with School of Electrical, Computer and Energy Engineering, Arizona State University, Tempe, USA. Email: xchen179@asu.edu. The paper has been accepted by IEEE Transactions on Parallel and Distributed Systems.}
}

\IEEEcompsoctitleabstractindextext{
\begin{abstract}
Mobile cloud computing is envisioned as a promising approach to augment computation capabilities of mobile devices for emerging resource-hungry mobile applications. In this paper, we propose a game theoretic approach for achieving efficient computation offloading for mobile cloud computing. We formulate the decentralized computation offloading decision making problem among mobile device users as a decentralized computation offloading game. We analyze the structural property of the game and show that the game always admits a Nash equilibrium. We then design a decentralized computation offloading mechanism that can achieve a Nash equilibrium of the game and quantify its efficiency ratio over the centralized optimal solution. Numerical results demonstrate that the proposed mechanism can achieve efficient computation offloading performance and scale well as the system size increases.

\end{abstract}

%\IEEEpeerreviewmaketitle
}

\maketitle
\pagestyle{empty}
\thispagestyle{empty}

\allowdisplaybreaks

\section{Introduction}

As smart-phones are gaining enormous popularity, more and more new
mobile applications such as face recognition, natural language processing,
interactive gaming, and augmented reality are emerging and attract
great attention \cite{soyata2012cloud,cohen2008embedded}. This kind of mobile applications
are typically resource-hungry, demanding intensive computation and
high energy consumption. Due to the physical size constraint, however,
mobile devices are in general resource-constrained, having limited
computation resources and limited battery life. The tension between
resource-hungry applications and resource-constrained mobile devices
hence poses a significant challenge for the future mobile platform development \cite{cuervo2010maui}.

\begin{figure}[t]
\centering
\includegraphics[scale=0.5]{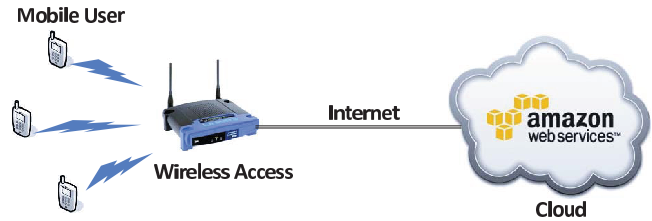}
\caption{\label{fig:An-illustration-of}An illustration of mobile cloud computing }
\end{figure}

Mobile cloud computing is envisioned as a promising approach to address
such a challenge. As illustrated in Figure \ref{fig:An-illustration-of}, mobile cloud computing can augment the capabilities
of mobile devices for resource-hungry applications, by offloading the computation  via wireless access to
the resource-rich cloud infrastructure such as Amazon Elastic Compute Cloud (EC2) and Windows
Azure Services Platform. In the cloud, each mobile device is associated with a cloud clone, which runs on a virtual
machine (VM) that can execute mobile applications on behalf of the mobile device\footnote{In this study we focus on the mobile application services (e.g., remote application execution) of the cloud. However, the cloud can also provide a number of other services  \cite{bahl2012advancing}, such as platform services (e.g., storage and file backup services). } \cite{chun2011clonecloud,wen2012energy}.

%However, an evident weakness of public cloud
%based mobile cloud computing is that mobile users may experience long
%latency for the data exchange with the public cloud in the wide area
%network. Long latency would hurt the interactive response, since humans
%are acutely sensitive to delay and jitter. Moreover, it is very difficult
%to control the latency in the wide area network. To overcome this
%limitation, another promising approach, \emph{cloudlet} based mobile
%cloud computing, is proposed in \cite{satyanarayanan2009case}. Rather than relying
%on a remote cloud, cloudlet based mobile cloud computing leverages
%the physical proximity and offloads the computation to the nearby
%cloud infrastructure (e.g., a cluster of multi-core computers), called
%cloudlet \cite{satyanarayanan2009case}. In this case, the need for fast interactive
%response can be met by low-latency, one-hop wireless access to the
%cloudlet. The cloudlets are envisioned to be deployed much like the
%WiFi service today as a computation augmenting service for mobile
%phone users on business premises such as a coffee shop or a doctor's office \cite{satyanarayanan2009case}.

Although the cloud based approach can significantly augment computation capability of mobile device users,
the task of developing a comprehensive and reliable mobile cloud computing
system remains challenging. A key challenge is how to achieve an efficient
computation offloading coordination among mobile device users. One critical factor of affecting the performance of mobile cloud computing is the wireless access efficiency \cite{barbera2013offload}. If too many mobile device users choose to offload the computation to the cloud via wireless access
simultaneously, they may generate severe interference to each other, which would reduce the data rates for computation
data transmission. This hence can lead to low energy efficiency for
computation offloading and long data transmission time. In this case,
it would not be beneficial for the mobile device users to offload computation
to the cloud.

In this paper, we adopt a game theoretic approach to address such a
challenge. Game theory is a useful framework for designing decentralized
mechanisms, such that the mobile device users in the system can self-organize
into the mutually satisfactory computation offloading decisions. The
self-organizing feature can add autonomics into mobile cloud computing system and help to ease the heavy burden of complex centralized
management (e.g.,  information collection from massive mobile device users and
computation offloading scheduling) by the cloud. Moreover, as different mobile devices are
usually owned by different individuals and they may pursue
different interests, game theory is a powerful tool to analyze
the interactions among multiple mobile device users who act in their own interests and devise incentive compatible computation offloading mechanisms such that no mobile user has the incentive to deviate unilaterally.

Specifically, we model the decentralized computation offloading decision making
problem among mobile device users for mobile cloud computing
as a decentralized computation offloading game. We then propose a
decentralized computation offloading mechanism that can achieve the
Nash equilibrium of the game. The main results and contributions of
this paper are as follows:
\begin{itemize}
\item \emph{Decentralized computation offloading game formulation}: We formulate
the decentralized computation offloading decision making problem among multiple mobile device users as a decentralized computation offloading game, by taking into account both communication and computation aspects of mobile cloud computing.
\item \emph{Analysis of game structure}: We analyze the decentralized computation offloading game
in both homogenous and heterogeneous wireless access cases. For
the homogenous case, we show that the game admits the beneficial cloud
computing group structure, which guarantees the existence
of Nash equilibrium. For the more general heterogeneous case, we show that the
game is a potential game, and hence admits the finite improvement property and possesses a Nash equilibrium.
\item \emph{Decentralized mechanism for achieving Nash equilibrium}: We
devise a decentralized computation offloading mechanism such that mobile device users make decisions locally, which can significantly
reduce the controlling and signaling overhead of the cloud. We
show that the mechanism can achieve a Nash equilibrium of the decentralized
computation offloading game. We further quantify the price of anarchy, i.e.,
the efficiency ratio of the mechanism over the centralized optimal solution. Numerical results demonstrate that the proposed mechanism can achieve efficient computation offloading performance and scale well as the system size increases.
\end{itemize}

The rest of the paper is organized as follows. We first discuss related
work in Section \ref{sec:Related-Work}, and introduce the system model
in Section \ref{sec:System-Model}. We then propose the decentralized
computation offloading game and develop the decentralized computation
offloading mechanism in Sections \ref{sec:Decentralized-Computation-Offloa}
and \ref{sec:Decentralized-Computation-Offloa-1}, respectively. We
present the numerical results in Section \ref{sec:Numerical-Results}
and finally conclude in Section \ref{sec:Conclusion}.

\section{Related Work}\label{sec:Related-Work}
Most previous work has investigated the efficient computation offloading mechanism design from the perspective of a single mobile device user \cite{kumar2013survey,rahimi2013mobile,rahimi2013mobile,kumar2010cloud,huang2012dynamic,rudenko1998saving,huerta2008adaptable,xian2007adaptive,barbera2013offload,wolski2008using,wen2012energy}.  Rudenko \emph{et al.} in \cite{rudenko1998saving} demonstrated by experiments that significant energy can be saved by computation offloading.  Gonzalo \emph{et al.}  in \cite{huerta2008adaptable} developed an adaptive offloading algorithm based on both the execution history of applications and the current system conditions. Xian \emph{et al.}  in \cite{xian2007adaptive} introduced an efficient timeout scheme for computation offloading to increase the energy efficiency on mobile devices. Rahimi \emph{et al.} in \cite{rahimi2012mapcloud} proposed a 2-tier cloud architecture to improve both performance and scalability of mobile cloud computing. Huang \emph{et al.} in \cite{huang2012dynamic} proposed a Lyapunov optimization based dynamic offloading algorithm to improve the mobile cloud computing performance while meeting the application execution time. Barbera \emph{et al.} in \cite{barbera2013offload} showed by realistic measurements that the wireless access plays a key role in affecting the performance of mobile cloud computing. Wolski \emph{et al.} in  \cite{wolski2008using} proposed a prediction based decision making framework for determining when an offloaded computation will outperform local execution on the mobile device. Wen \emph{et al.} in \cite{wen2012energy} presented an efficient offloading policy by jointly configuring the clock frequency in the mobile device and scheduling the data transmission to minimize the energy consumption.

To the best of our knowledge, only a few works have addressed the computation offloading problem under the setting of multiple mobile device users  \cite{yang2013framework,rahimimusic2013,barbarossa2013joint}.  Yang \emph{et al.} in \cite{yang2013framework} studied the scenario that multiple users share the wireless network bandwidth, and solved the problem of maximizing the mobile cloud computing performance by a centralized heuristic genetic algorithm. Rahimi \emph{et al.} in \cite{rahimimusic2013} took into consideration user mobility information and proposed a centralized greedy scheme to solve the computation offloading problem with multiple mobile users. Barbarossa \emph{et al.} in \cite{barbarossa2013joint} proposed a centralized scheduling algorithm to jointly optimize the communication and computation resource allocations among multiple users with the latency requirements. The centralized computation offloading schemes above requires that all the mobile device users submit their own information (e.g., wireless channel gain and the size of computation tasks) to a centralized entity (e.g., the cloud), which will determine the offloading schedule accordingly. Along a different line, in this paper we adopt the game theoretic approach and devise a decentralized mechanism wherein each mobile device user makes the computation offloading decision locally. This can help to reduce the controlling and signaling overhead of the cloud.

\section{System Model\label{sec:System-Model}}
In this section, we introduce the system model of mobile cloud computing. We consider a set of $\mathcal{N}=\{1,2,...,N\}$ collocated mobile device users and each of which has a computationally intensive and delay sensitive task to be completed. There exists a wireless access base-station $s$, through which the mobile device users can offload the computation to the cloud (e.g., Amazon EC2 or Microsoft Azure). Similar to many previous studies in mobile cloud computing \cite{yang2013framework,wen2012energy,barbarossa2013joint} and mobile networking \cite{wu2002multi,iosifidisiterative2013,wu2005minimum}, to enable tractable analysis and get useful insights, we consider a quasi-static scenario where the set of mobile device users $\mathcal{N}$ remains unchanged  during a computation offloading period (e.g., within several seconds), while may change across different periods\footnote{This assumption holds for many applications such as face recognition and natural language processing, in which the size of computation input data is not large and hence the computation offloading can be finished in a smaller time scale (e.g., within several seconds) than the time scale of users' mobility.}. The general case that mobile users may depart and leave dynamically within a computation offloading period will be considered in a future work. Since both the communication and computation aspects play a key role in mobile cloud computing, we next introduce the communication and computation models in details.

\subsection{Communication Model \label{sub:Communication-Model}}

We first introduce the communication model for wireless access. The wireless access base-station $s$ can be either a WiFi access point, or a Femtocell network access point \cite{lopez2013distributed}, or a macrocell base-station in cellular networks that manages the uplink/downlink communications of mobile device users. We denote $a_{n}\in\{0,1\}$ as the computation offloading decision of mobile device user $n$. Specifically, we have $a_{n}=1$ if user $n$ chooses to offload the computation to the cloud via wireless access. We have $a_{n}=0$ if user $n$ decides to compute its task locally on the mobile device.
Given the decision profile $\boldsymbol{a}=(a_{1},a_{2},...,a_{N})$
of all the mobile device users, we can compute the uplink data rate for computation offloading of mobile device user $n$ as \cite{rappaport1996wireless}
\begin{equation}
R_{n}(\boldsymbol{a})=W\log_{2}\left(1+\frac{P_{n}H_{n,s}}{\omega_{n}+\sum_{m\in\mathcal{N}\backslash\{n\}:a_{m}=1}P_{m}H_{m,s}}\right).\label{eq:R1}
\end{equation}
Here $W$ is the channel bandwidth and $P_{n}$ is user $n$'s transmission
power which is determined by the wireless access base-station according to some power
control algorithms such as \cite{xiao2003utility,saraydar2002efficient}. Further, $H_{n,s}$ denotes
the channel gain between the mobile device user $n$ and the base-station,
and $\omega_{n}=\omega_{n}^{0}+\omega_{n}^{1}$ denotes the background
interference power including the noise power $\omega_{n}^{0}$ and
the interference power $\omega_{n}^{1}$ from other mobile device users who carry out wireless transmission but do not involve in the mobile cloud computing.

From the communication model in (\ref{eq:R1}), we see that if too many mobile device users  choose to offload the computation via wireless access simultaneously, they may incur severe interference, leading to low data rates. As we discuss latter, this would negatively affect the performance of mobile cloud computing.

\subsection{Computation Model}

We then introduce the computation model. We consider that each mobile device
user $n$ has a computation task $\mathcal{I}_{n}\triangleq(B_{n},D_{n})$
that can be computed  either locally on the mobile device or remotely
on the cloud via computation offloading. Here $B_{n}$ denotes the
size of computation input data (e.g., the program codes and input parameters) involving in the computation task $\mathcal{I}_{n}$
and $D_{n}$ denotes the total number of CPU cycles required to accomplish the computation task $\mathcal{I}_{n}$. A mobile device user $n$ can apply the methods in \cite{cuervo2010maui,yang2013framework,chun2011clonecloud} to obtain the information of $B_{n}$ and $D_{n}$.   We
next discuss the computation overhead in terms of both energy consumption and processing time for both local and cloud computing approaches.

\subsubsection{Local Computing}

For the local computing approach, a mobile device user $n$ executes its
computation task $\mathcal{I}_{n}$ locally on the mobile device. Let
$F^{l}_{n}$ be the computation capability (i.e., CPU cycles per
second) of mobile device user $n$. Here we allow that different mobile devices may have different computation capability. The computation execution time of
the task $\mathcal{I}_{n}$ by local computing is then given as
\begin{equation}
T_{n}^{l}=\frac{D_{n}}{F^{l}_{n}}.\label{eq:l1}
\end{equation}
For the computational energy, we have that
\begin{equation}
E_{n}^{l}=\nu_{n}D_{n},\label{eq:l2}
\end{equation}
where $\nu_{n}$ is the coefficient denoting the consumed energy per CPU cycle. According
to the realistic measurements in \cite{miettinen2010energy,wen2012energy}, we can set $\nu_{n}=10^{-11} (F^{l}_{n})^{2}$.

According to (\ref{eq:l1}) and (\ref{eq:l2}), we can then compute
the overhead of the local computing approach in terms of computational
time and energy as
\begin{equation}
Z_{n}^{l}=\gamma_{n}^{T}T_{n}^{l}+\gamma_{n}^{E}E_{n}^{l},\label{eq:l3}
\end{equation}
where $0\leq\gamma_{n}^{T},\gamma_{n}^{E}\leq1$ denote the weights of
computational time and energy for mobile device user $n$'s decision making,
respectively. To provide rich modeling flexibility and meet user-specific demands, we allow that different users can choose different weighting parameters in the decision making. For example, when a user is at a low battery state, the user would like to put more weight on energy consumption (i.e., a larger $\gamma_{n}^{E}$) in the decision making, in order to save more energy. When a user is running some application that is sensitive to the delay (e.g., video streaming), then the user can put more weight on the processing time (i.e., a larger $\gamma_{n}^{T}$), in order to reduce the delay. Note that the weights could be dynamic if a user runs different applications or has different policies/demands at different computation offloading periods. For ease of exposition, in this paper we assume that the weights of a user are fixed within one computation offloading period, while can be changed in different periods.

\subsubsection{Cloud Computing}

For the cloud computing approach, a mobile device user $n$ will offload
its computation task $\mathcal{I}_{n}$ to the cloud and the cloud
will execute the computation task on behalf of the mobile device user.

For the computation offloading, a mobile device user $n$ would incur the extra
overhead in terms of time and energy for transmitting the computation input data
to the cloud via wireless access. According to the communication
model in Section \ref{sub:Communication-Model}, we can compute the
transmission time and energy of mobile device user $n$ for offloading the input
data of size $B_{n}$ as, respectively,
\begin{equation}
T_{n,off}^{c}(\boldsymbol{a})=\frac{B_{n}}{R_{n}(\boldsymbol{a})},\label{eq:c2}
\end{equation}
and
\begin{equation}
E_{n}^{c}(\boldsymbol{a})=\frac{P_{n}B_{n}}{R_{n}(\boldsymbol{a})}.\label{eq:c3}
\end{equation}

After the offloading, the cloud will execute the computation
task $\mathcal{I}_{n}$. Let $F^{c}_{n}$ be the computation capability
(i.e., CPU cycles per second) assigned to user $n$ by the cloud. The execution time
of the task $\mathcal{I}_{n}$ of mobile device user $n$ on the cloud can
be then given as
\begin{equation}
T_{n,exe}^{c}=\frac{D_{n}}{F^{c}_{n}}.\label{eq:c1}
\end{equation}

According to (\ref{eq:c2}), (\ref{eq:c3}), and (\ref{eq:c1}), we
can compute the overhead of the cloud computing approach in terms
of processing time and energy as
\begin{equation}
Z_{n}^{c}(\boldsymbol{a})=\gamma_{n}^{T}\left(T_{n,off}^{c}(\boldsymbol{a})+T_{n,exe}^{c}\right)+\gamma_{n}^{E}E_{n}^{c}(\boldsymbol{a}).\label{eq:c4}
\end{equation}

Similar to many studies such as \cite{kumar2010cloud,huang2012dynamic,rudenko1998saving,huerta2008adaptable,xian2007adaptive}, we neglect
the time overhead for the cloud to send the computation outcome back
to the mobile device user, due to the fact that for many applications (e.g., face recognition), the size of the computation outcome in
general is much smaller than the size of computation input data including the mobile system settings, program codes and input parameters.

According to the communication and computation models above, we see
that the computation offloading decisions $\boldsymbol{a}$ among the mobile device users
are coupled. If too many mobile device users simultaneously choose to offload the computation
task to the cloud via wireless access, they may incur severe
interference and  this would lead to a low data rate. When the data rate $R_{n}(\boldsymbol{a})$
of a mobile device user $n$ is low, it would consume high energy in the
wireless access for offloading the computation input data to
cloud and incur long transmission time as well. In
this case, it would be more beneficial for the user to compute the
task locally on the mobile device to avoid the long processing time and high energy consumption by the cloud computing approach. In the following sections, we will
adopt a game theoretic approach to address the issue of how to achieve
efficient computation offloading decision makings among the mobile device users.

\section{Decentralized Computation Offloading Game\label{sec:Decentralized-Computation-Offloa}}

In this section, we develop a game theoretic approach for achieving
efficient computation offloading decision makings among the mobile device
users. The primary rationale of adopting the game theoretic approach is that
the mobile devices are owned by different individuals and they may
pursue different interests. Game theory is a powerful framework to analyze
the interactions among multiple mobile device users who act in their own interests and devise incentive compatible computation offloading mechanisms such that no user has the incentive to deviate unilaterally. Moreover, by leveraging the intelligence of each individual mobile device user, game theory is a useful tool for devising decentralized mechanisms with low complexity, such that the users can self-organize into a mutually satisfactory solution. This can help to ease the heavy burden of complex centralized management by the cloud and reduce the controlling and signaling overhead between the cloud and mobile device users.

\subsection{Game Formulation}

We consider the decentralized computation offloading decision making problem among
the mobile device users within a computation offloading period. Let $a_{-n}=(a_{1},...,a_{n-1},a_{n+1},...,a_{N})$
be computation offloading decisions by all other users except
user $n$. Given other users' decisions $a_{-n}$, user $n$
would like to select a proper decision $a_{n}\in\{0,1\}$ (i.e., local
computing or cloud computing) to minimize its computation overhead in
terms of energy consumption and processing time, i.e.,
\[
\min_{a_{n}\in\{0,1\}}V_{n}(a_{n},a_{-n}),\forall n\in\mathcal{N}.
\]
 According to (\ref{eq:l3}) and (\ref{eq:c4}), we can obtain the
overhead function of mobile device user $n$ as
\begin{equation}
V_{n}(a_{n},a_{-n})=\begin{cases}
Z_{n}^{l}, & \mbox{if }a_{n}=0,\\
Z_{n}^{c}(\boldsymbol{a}), & \mbox{if }a_{n}=1.
\end{cases}\label{eq:V1}
\end{equation}

We then formulate the problem above as a strategic game $\Gamma=(\mathcal{N},\{\mathcal{A}_{n}\}_{n\in\mathcal{N}},\{V_{n}\}_{n\in\mathcal{N}})$,
where the set of mobile device users $\mathcal{N}$ is the set of players,
$\mathcal{A}_{n}\triangleq\{0,1\}$ is the set of strategies for user
$n$, and the overhead function $V_{n}(a_{n},a_{-n})$ of each user
$n$ is the cost function to be minimized by player $n$. In the sequel,
we call the game $\Gamma$ as the decentralized computation offloading game.
We now introduce the concept of Nash equilibrium \cite{osborne1994course}.
\begin{defn}
A strategy profile $\boldsymbol{a}^{*}=(a_{1}^{*},...,a_{N}^{*})$
is a Nash equilibrium of the decentralized computation offloading
game if at the equilibrium $\boldsymbol{a}^{*}$, no player can further reduce its overhead by unilaterally
changing its strategy, i.e.,
\begin{equation}
V_{n}(a_{n}^{*},a_{-n}^{*})\leq V_{n}(a_{n},a_{-n}^{*}),\forall a_{n}\in\mathcal{A}_{n},n\in\mathcal{N}.\label{eq:ne1}
\end{equation}
\end{defn}

The Nash equilibrium has the nice self-stability property such that
the users at the equilibrium can achieve a mutually satisfactory solution and no user has the incentive to deviate. This property is very important to the decentralized computation offloading problem, since the mobile devices are owned by different individuals and they may act in their own interests.

\subsection{Game Property}

We then study the existence of Nash equilibrium of the decentralized
computation offloading game. To proceed, we first introduce an important
concept of best response \cite{osborne1994course}.
\begin{defn}
\label{Given-the-strategies}Given the strategies $a_{-n}$ of the other
players, player $n$'s strategy $a_{n}^{*}\in\mathcal{A}_{n}$ is
a best response if
\begin{equation}
V_{n}(a_{n}^{*},a_{-n})\leq V_{n}(a_{n},a_{-n}),\forall a_{n}\in\mathcal{A}_{n}.\label{eq:bs1}
\end{equation}
\end{defn}

According to (\ref{eq:ne1}) and (\ref{eq:bs1}), we see that at the
Nash equilibrium all the users play the best response strategies towards
each other. Based on the concept of best response, we have the following
observation for the decentralized computation offloading game.
\begin{lem}
\label{lem:Given-the-strategies}Given the strategies $a_{-n}$ of
other mobile device users in the decentralized computation offloading game,
the best response of a user $n$ is given as the following threshold
strategy
\[
a_{n}^{*}=\begin{cases}
1, & \mbox{if }\sum_{m\in\mathcal{N}\backslash\{n\}:a_{m}=1}P_{m}H_{m,s}\leq L_{n},\\
0, & \mbox{otherwise,}
\end{cases}
\]
where the threshold
\[
L_{n}=\frac{P_{n}H_{n,s}}{2^{\frac{\left(\gamma_{n}^{T}+\gamma_{n}^{E}P_{n}\right)B_{n}}{W\left(\gamma_{n}^{T}T_{n}^{l}+\gamma_{n}^{E}E_{n}^{l}-\gamma_{n}^{T}T_{n,exe}^{c}\right)}}-1}-\omega_{n}.
\]
\end{lem}

The proof is given in Section 8.1 of the separate supplementary file. According to Lemma \ref{lem:Given-the-strategies}, we see that when
the received interference $\sum_{m\in\mathcal{N}\backslash\{n\}:a_{m}=1}P_{m}H_{m,s}$
is lower enough, it is beneficial for user $n$ to offload the computation
to the cloud. Otherwise, the user $n$ should compute the task on
the mobile device locally. Since the wireless access plays a
critical role in mobile cloud computing, we next discuss the existence of Nash equilibrium
of the the decentralized computation offloading game in both homogeneous
and heterogeneous wireless access cases.

\subsubsection{Homogeneous Wireless Access Case}

We first consider the case that users' wireless access is homogenous,
i.e., $P_{m}H_{m,s}=P_{n}H_{n,s}=K,$ for any $n,m\in\mathcal{N}$.
This can correspond to the scenario that all the mobile device users experience the similar channel condition and are assigned
with the same transmission power by the base-station. However, different
users may have different thresholds $L_{n}$, i.e., they are
heterogeneous in terms of computation capabilities and tasks.

\begin{algorithm}[tt]
\begin{algorithmic}[1]
\State \textbf{Input}: the set of ordered mobile device users with $\frac{L_{1}}{K}\geq\frac{L_{2}}{K}\geq...\geq\frac{L_{N}}{K}$ and $\frac{L_{1}}{K}\geq0$.
\State \textbf{Output}: a beneficial cloud computing group $\mathcal{S}$.
\State \textbf{set} $\mathcal{S}=\{1\}$.
\For{$t=2$ to $N$}
    \State \textbf{set} $\tilde{\mathcal{S}}=\mathcal{S}\cup\{t\}$
    \If{$|\tilde{\mathcal{S}}|>\frac{L_{t}}{K}+1$}
        \State \textbf{stop} and go to \textbf{return}.
    \Else{ \textbf{set} $\mathcal{S}=\tilde{\mathcal{S}}.$}
    \EndIf
\EndFor
\State \textbf{return} $\mathcal{S}$.

\end{algorithmic}
\caption{\label{alg:Beneficial-cloud-computing} Algorithm for finding beneficial cloud computing group}
\end{algorithm}

For the homogenous wireless access case, without loss of generality,
we can order the set $\mathcal{N}$ of mobile device users so that $\frac{L_{1}}{K}\geq\frac{L_{2}}{K}\geq...\geq\frac{L_{N}}{K}$.
Based on this, we have the following useful observation.
\begin{lem}\label{lem_homo}
For the decentralized computation offloading game with homogenous
wireless access, if there exists a non-empty beneficial cloud computing
group of mobile device users $\mathcal{S}\subseteq\mathcal{N}$ such that
\begin{eqnarray}
|\mathcal{S}| & \leq & \frac{L_{i}}{K}+1,\forall i\in\mathcal{S},\label{eq:h1}
\end{eqnarray}
and further if $\mathcal{S}\subset\mathcal{N}$,
\begin{equation}
|\mathcal{S}|>\frac{L_{j}}{K},\forall j\in\mathcal{N}\backslash\mathcal{S},\label{eq:h2}
\end{equation}
then the strategy profile wherein users $i\in\mathcal{S}$ play the
strategy $a_{i}=1$ and the other users $j\in\mathcal{N}\backslash\mathcal{S}$
play the strategy $a_{j}=0$ is a Nash equilibrium. \end{lem}

The proof is given in Section 8.2 of  the separate supplementary file. For example, for a set of $4$ users with
$\left(\frac{L_{1}}{K},\frac{L_{2}}{K},\frac{L_{3}}{K},\frac{L_{4}}{K}\right)=(5,4,3,2)$,
the beneficial cloud computing group is $\mathcal{S}=\{1,2,3\}$. In general, when $\frac{L_{1}}{K}\geq0$, we can construct the beneficial cloud computing group
by using Algorithm \ref{alg:Beneficial-cloud-computing}. Thus, we have
the following result.
\begin{thm}\label{thm:homo}
The decentralized computation offloading game with homogenous wireless access always has a Nash equilibrium. More specifically, when
$\frac{L_{1}}{K}<0$, all users $n\in\mathcal{N}$ playing the strategy
$a_{n}=0$ is a Nash equilibrium. When $\frac{L_{1}}{K}\geq0$, we
can construct a beneficial cloud computing group $\mathcal{S}\neq\varnothing$
by Algorithm \ref{alg:Beneficial-cloud-computing} such that the strategy
profile wherein users $i\in\mathcal{S}$ play the strategy $a_{i}=1$
and the other users $j\in\mathcal{N}\backslash\mathcal{S}$ play the
strategy $a_{j}=0$ is a Nash equilibrium.
\end{thm}

The proof is given in Section 8.3 of  the separate supplementary file. Since the computational complexity of ordering operation (e.g., quicksort algorithm) is typically $\mathcal{O}(N\log N)$ and the construction procedure in Algorithm \ref{alg:Beneficial-cloud-computing} involves at most $N$ operations (with each operation of the complexity of $\mathcal{O}(1)$), the beneficial cloud computing group construction algorithm has a low computational complexity of  $\mathcal{O}(N\log N)$.  This implies that we can compute the Nash equilibrium of the decentralized computation offloading game in the homogenous wireless access case in a fast manner.

\subsubsection{General Wireless Access Case}

We next consider the general case including the case that users' wireless access can
be heterogeneous, i.e., $P_{m}H_{m,s}\neq P_{n}H_{n,s}$. Since mobile device
users may have different transmission power $P_{n}$, channel gain $H_{n,s}$ and
thresholds $L_{n}$, the analysis based on the beneficial cloud computing
group in the homogenous case can not apply here. We hence resort to
a power tool of potential game \cite{monderer1996potential}.
\begin{defn}
A game is called a potential game if it admits a potential function
$\Phi(\boldsymbol{a})$ such that for every $n\in\mathcal{N}$, $a_{-n}\in\prod_{i\neq n}\mathcal{A}_{i}$,
and $a_{n}^{'},a_{n}\in\mathcal{A}_{n}$, if
\begin{equation}
V_{n}(a_{n}^{'},a_{-n})<V_{n}(a_{n},a_{-n}),\label{eq:p1}
\end{equation}
we have
\begin{equation}
\Phi(a_{n}^{'},a_{-n})<\Phi(a_{n},a_{-n}).\label{eq:p2}
\end{equation}
\end{defn}
\begin{defn}
The event where a player $n$ changes to an action $a_{n}^{'}$ from
the action $a_{n}$ is a better response update if and only if its
cost function is decreased, i.e.,
\begin{equation}
V_{n}(a_{n}^{'},a_{-n})<V_{n}(a_{n},a_{-n}).\label{eq:p3}
\end{equation}
\end{defn}

An appealing property of the potential game is that it admits the
finite improvement property, such that any asynchronous better response
update process (i.e., no more than one player updates the strategy
at any given time) must be finite and leads to a Nash equilibrium \cite{monderer1996potential}. Here the potential function to a game has the same spirit as the Lyapunov function to a dynamical system. If a dynamic system is shown to have a Lyapunov function, then the system has a stable point. Similarly, if a game admits a potential function, the game must have a Nash equilibrium.

We now prove the existence of Nash equilibrium of the general decentralized
computation offloading game by showing that the game is a potential
game. Specifically, we define the potential function as
\begin{align}
\Phi(\boldsymbol{a}) = & \frac{1}{2}\sum_{n=1}^{N}\sum_{m\ne n}P_{n}H_{n,s}P_{m}H_{m,s}I_{\{a_{n}=1\}}I_{\{a_{m}=1\}} \nonumber \\
& +\sum_{n=1}^{N}P_{n}H_{n,s}L_{n}I_{\{a_{n}=0\}},\label{eq:p4}
\end{align}
where $I_{\{A\}}$ is the indicator function such as $I_{\{A\}}=1$
if the event $A$ is true and $I_{\{A\}}=0$ otherwise.
\begin{thm}
\label{thm:The-general-decentralized}The general decentralized computation
offloading game is a potential game with the potential function
as given in (\ref{eq:p4}), and hence always has a Nash equilibrium
and the finite improvement property.\end{thm}

The proof is given in Section 8.4 of  the separate supplementary file. Theorem \ref{thm:The-general-decentralized} implies that any asynchronous
better response update process is guaranteed to reach a Nash equilibrium within
a finite number of iterations. This motivates the algorithm design
in following Section \ref{sec:Decentralized-Computation-Offloa-1}.

\section{Decentralized Computation Offloading Mechanism\label{sec:Decentralized-Computation-Offloa-1}}

In this section we propose a decentralized computation offloading
 mechanism in Algorithm \ref{alg:Decentralized-computation-offloa}
for achieving the Nash equilibrium of the decentralized computation
offloading game.

\subsection{Mechanism Design}

The motivation of using the decentralized computation offloading mechanism is to coordinate mobile device
users to achieve a mutually satisfactory decision making, prior to the computation task execution. The key idea of the mechanism design is to utilize the
finite improvement property of the decentralized computation offloading game and let one mobile device user improve
its computation offloading decision at a time. Specifically, by using the clock signal from the wireless access base-station for synchronization, we consider a slotted time structure for the computation offloading decision update. Each decision slot $t$ consists the following two parts:
\begin{itemize}
\item \textbf{Interference Measurement}: Each mobile device user $n$ locally measures
the received interference $\mu_{n}(t)=\sum_{m\in\mathcal{N}\backslash\{n\}:a_{m}(t)=1}P_{m}H_{m,s}$
generated by other users who currently choose the decisions of
offloading the computation tasks to the cloud via wireless access. To facilitate the interference measurement,
for example, the users $m$ who choose decisions $a_{m}(t)=1$ at the current slot will transmit some pilot signals to the base-station. And each mobile device user can then enquire its received interference $\mu_{n}(t)$ from the base-station.
\item \textbf{Decision Update Contention}: We exploit the finite improvement
property of the game by having one mobile device user carry out a decision update at
each decision slot. We let users who can improve
their computation performance compete for the decision update opportunity
in a decentralized manner. More specifically, according to Lemma \ref{lem:Given-the-strategies}, each mobile device user $n$
first computes its set of best response update based on the measured interference $\mu_{n}(t)$ as
\begin{align*}
\Delta_{n}(t) & \triangleq \{  a_{n}^{*}: V_{n}(a_{n}^{*},a_{-n}(t)) <V_{n}(a_{n}(t),a_{-n}(t))\} \\
& =\begin{cases}
\{1\}, & \mbox{if }a_{n}(t)=0\mbox{ and }\mu_{n}(t)\leq L_{n},\\
\{0\}, & \mbox{if }a_{n}(t)=1\mbox{ and }\mu_{n}(t)>L_{n},\\
\varnothing, & \mbox{otherwise.}
\end{cases}
\end{align*}
The best response here is similar to the steepest descent direction selection to reduce user's overhead.   Then, if $\Delta_{n}(t)\neq\varnothing$ (i.e., user $n$ can improve),
user $n$ will contend for the decision update opportunity.
Otherwise, user $n$ will not contend and  adhere to the current
decision at next decision slot, i.e., $a_{n}(t+1)=a_{n}(t)$. For
the decision update contention, for example, we can adopt the random backoff-based
mechanism by setting the time length of decision update contention
as $\tau^{*}$. Each contending user $n$ first generates a backoff
time value $\tau_{n}$ according to the uniform distribution over $[0,\tau^{*}]$
and countdown until the backoff timer expires. When the timer expires,
if the user has not received any request-to-update (RTU) message from other mobile device users yet, the user will update
its decision for the next slot as $a_{n}(t+1)\in\Delta_{n}(t)$ and
then broadcast a RTU message to all users to indicate that it wins the decision update contention. For other users, on hearing the RTU message, they will not update their decisions and will choose the same decisions at next slot, i.e., $a_{n}(t+1)=a_{n}(t)$.
\end{itemize}

According to the finite improvement property in Theorem \ref{thm:The-general-decentralized},
the mechanism will converge to a Nash equilibrium of the decentralized
computation offloading game within finite number of decision slots. In practice, we can implement that the computation offloading decision update process terminates when no RTU messages are broadcasted for multiple consecutive decision slots (i.e., no decision update can be further carried out by any users). Then each mobile device user $n$ executes the computation task according to the decision $a_{n}$ obtained at the last decision slot by the mechanism. Due to the property of Nash equilibrium, no user has the incentive to deviate from the achieved decisions. This is very important to the decentralized computation offloading problem, since the mobile devices are owned by different individuals and they may act in their own interests. By following the decentralized computation offloading mechanism,  the users adopt the best response to improve their decision makings and eventually self-organize into a mutually satisfactory solution (i.e., Nash equilibrium).

We then analyze the computational complexity of the algorithm. In each iteration, $N$ mobile users will execute the operations in Lines $5-15$. Since the operations in Lines $5-15$ only involve some basic arithmetical calculations, the computational complexity in each iteration is $\mathcal{O}(N)$. Suppose that it takes $C$ iterations for the algorithm to converge. Then the total computational complexity of the algorithm is $\mathcal{O}(CN)$. Numerical results in Section \ref{sec:Numerical-Results} show that the number of iterations $C$ for convergence increases linearly with the number of users $N$. This demonstrates that the decentralized computation offloading mechanism can converge in a fast manner in practice.

\begin{algorithm}[tt]
\begin{algorithmic}[1]
\State \textbf{initialization:}
\State each mobile device user $n$ \textbf{chooses} the computation decision $a_{n}(0)=1$.
\State \textbf{end initialization\newline}

\Repeat{ for each user $n$ and each decision slot $t$ in parallel:}
        \State \textbf{measure}  the interference $\mu_{n}(t)$.
        \State \textbf{compute}  the best response set $\Delta_{n}(t)$.
        \If{ $\Delta_{n}(t)\neq\varnothing$}
            \State \textbf{contend}  for the decision update opportunity.
            \If{ \textbf{win} the decision update contention}
                \State \textbf{choose}  the decision $a_{n}(t+1)\in\Delta_{n}(t)$ for next slot.
                \State \textbf{broadcast} the RTU message to other users.
            \Else{ choose the original decision $a_{n}(t+1)=a_{n}(t)$ for next slot.}
            \EndIf
        \Else{ \textbf{choose} the original decision $a_{n}(t+1)=a_{n}(t)$ for next slot.}
        \EndIf
%\EndLoop
\Until{no RTU messages are broadcasted for $M$ consecutive slots}

\end{algorithmic}
\caption{\label{alg:Decentralized-computation-offloa}Decentralized computation offloading mechanism}
\end{algorithm}

%\begin{algorithm}
%Initialization: each mobile device user $n$ chooses computation decision
%$a_{n}(0)=1$.
%
%Repeat for each user $n$ and each decision slot $t$ do
%
%Measure the interference $\mu_{n}(t)$;
%
%Compute the better response set $\Delta_{n}(t)$;
%
%If $\Delta_{n}(t)\neq\varnothing$ then
%
%Contend for the decision update opportunity;
%
%if win the decision update contention then
%
%choose the decision $a_{n}(t+1)\in\Delta_{n}(t)$ for next slot.
%
%broadcast the RTS message for decision updating to other users.
%
%else choose the original decision $a_{n}(t+1)=a_{n}(t)$ for next slot
%
%Else, choose the original decision $a_{n}(t+1)=a_{n}(t)$ for next slot
%
%Until no RTS messages for decision updating are broadcasted for $H$
%consecutive slots
%
%\caption{\label{alg:Decentralized-computation-offloa}Decentralized computation
%offloading decision mechanism}
%\end{algorithm}

\subsection{Performance Analysis}

We then discuss the efficiency of Nash equilibrium by the decentralized
computation offloading mechanism. Note that the decentralized
computation offloading game may have multiple Nash equilibria,
and the proposed decentralized computation offloading mechanism
will randomly select one Nash equilibrium (since a random user is chosen for decision update). Following the definition of price
of anarchy (PoA) in game theory \cite{roughgarden2005selfish}, we will quantify the efficiency
ratio of the worst-case Nash equilibrium over the centralized optimal
solution. Let $\Upsilon$ be the set of Nash equilibria of the decentralized
computation offloading game. Then the PoA is defined as
\[
\mbox{PoA}=\frac{\max_{\boldsymbol{a}\in\Upsilon}\sum_{n\in\mathcal{N}}V_{n}(\boldsymbol{a})}{\min_{\boldsymbol{a}\in\prod_{n=1}^{N}\mathcal{A}_{n}}\sum_{n\in\mathcal{N}}V_{n}(\boldsymbol{a})},
\]
which is lower bounded by $1$. A larger PoA implies that the set
of Nash equilibrium is less efficient (in the worst-case sense) using
the centralized optimum as a benchmark. Let $\overline{Z_{n}^{c}}=\frac{\left(\gamma_{n}^{T}+\gamma_{n}^{E}P_{n}\right)B_{n}}{W\log_{2}\left(1+\frac{P_{n}H_{n,s}}{\omega_{n}}\right)}+\gamma_{n}^{T}T_{n,exe}^{c}.$
We can show the following result.
\begin{thm}
\label{thm:The-PoA-of}The PoA of the decentralized computation offloading
game is at most $\frac{\sum_{n=1}^{N}Z_{n}^{l}}{\sum_{n=1}^{N}\min\{Z_{n}^{l},\overline{Z_{n}^{c}}\}}.$\end{thm}

The proof is given in Section 8.5 of  the separate supplementary file. Intuitively, Theorem \ref{thm:The-PoA-of} indicates that when users
have lower cost of local computing (i.e., $Z_{n}^{l}$ is smaller),
the Nash equilibrium is closer to the centralized optimum and hence
the PoA is lower. Moreover, when the communication efficiency is higher
(i.e., $P_{n}H_{n,s}$ is larger and hence $\overline{Z_{n}^{c}}$
is larger), the performance of Nash equilibrium can be improved. Numerical results in Section \ref{sec:Numerical-Results} demonstrate
that the Nash equilibrium by the decentralized computation offloading
mechanism is efficient, with at most $10\%$ performance loss,
compared with the centralized optimal solution.

\section{Numerical Results\label{sec:Numerical-Results}}

%\begin{figure}
%\centering
%\includegraphics[scale=0.7]{Map}
%\caption{\label{fig:A-square-area}A square area of a length of $50$m with
%the wireless access base-station located in the center.}
%\end{figure}

\begin{figure}
\centering
\includegraphics[scale=0.4]{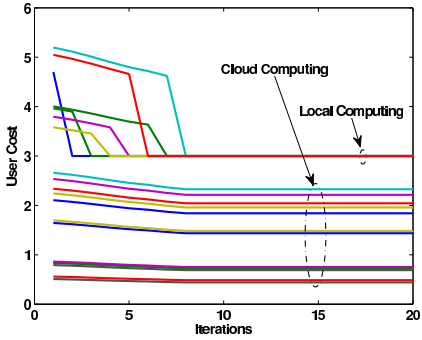}
\caption{\label{fig:Dynamics-of-user}Dynamics of user cost by the decentralized
computation offloading mechanism}
\end{figure}

In this section, we evaluate the proposed decentralized computation
offloading mechanism by numerical studies. We first consider the mobile
cloud computing scenario that $N=20$ mobile device users are randomly scattered
over a $50$m$\times$$50$m region and the wireless access base-station is located in
the center of the region. For the wireless access, we set the channel bandwidth
$W=5$ MHz, the transmission power $P_{n}=100$ mWatts, and the background
noise $\omega_{n}=-100$ dBm. According to the physical interference
model \cite{rappaport1996wireless}, we set the channel gain $H_{n,s}=d_{n,s}^{-\alpha}$,
where $d_{n,s}$ is the distance between mobile device user $n$ and the
cloudlet and $\alpha=4$ is the path loss factor.  We set the decision weights $\gamma_{n}^{T}=\gamma_{n}^{E}=0.5$. For the computation
task, we use the face recognition application in \cite{soyata2012cloud},
where the data size for the computation offloading $B_{n}=420$ KB
and the total number of CPU cycles $D_{n}=1000$ Megacycles.
The CPU computational capability $F^{l}_{n}$ of a mobile device
user $n$ is randomly assigned from the set $\{0.5,0.8,1.0\}$ GHz
and the computational capability on the cloud $F^{c}_{n}=100$ GHz \cite{soyata2012cloud}.

\begin{figure*}[tt]
\begin{minipage}[t]{0.32\linewidth}
\centering
\includegraphics[scale=0.4]{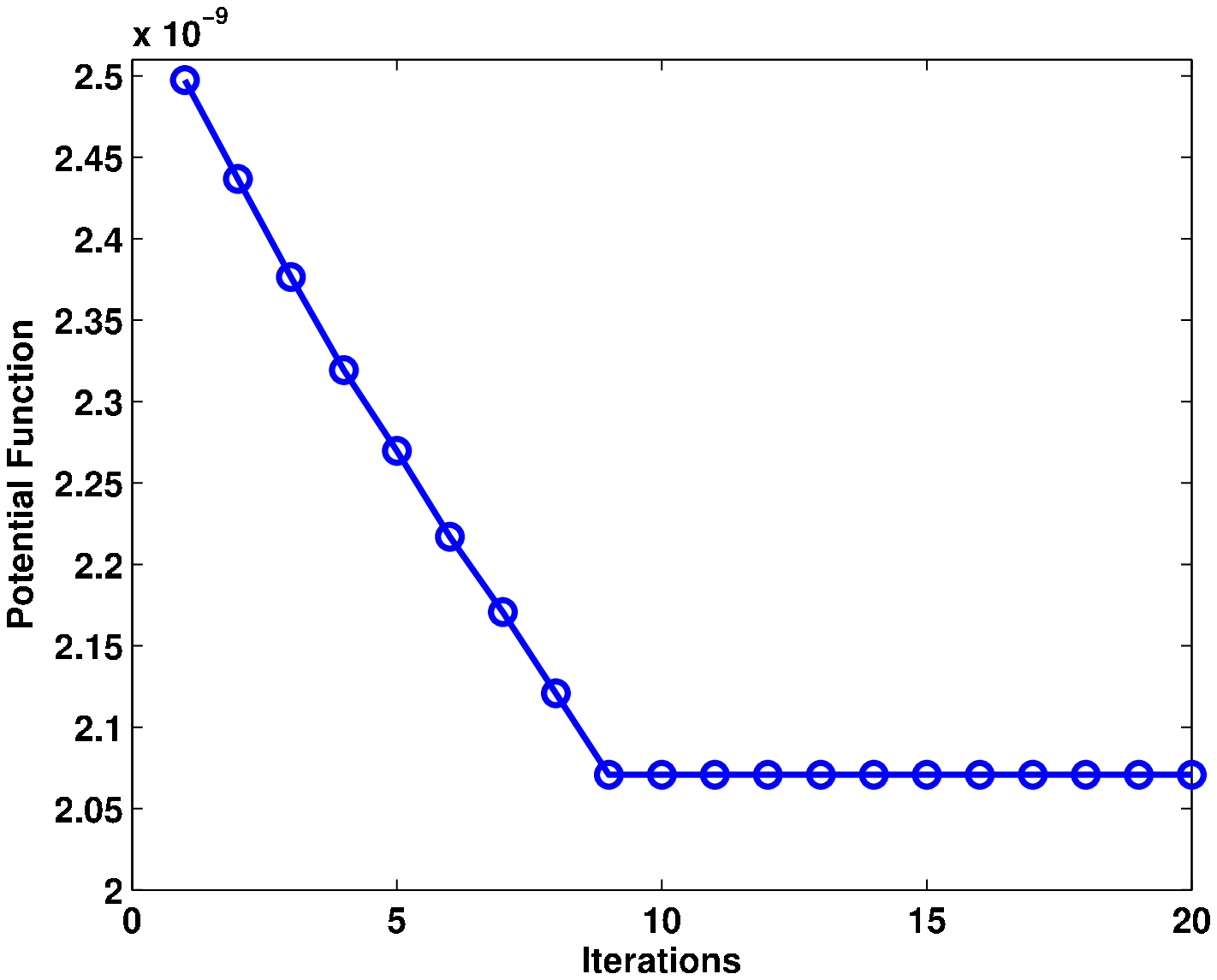}
\caption{Dynamics of potential function \label{fig:Dynamics-of-potential}by
the decentralized computation offloading mechanism}
\end{minipage}
\hfill
\begin{minipage}[t]{0.32\linewidth}
\centering
\includegraphics[scale=0.4]{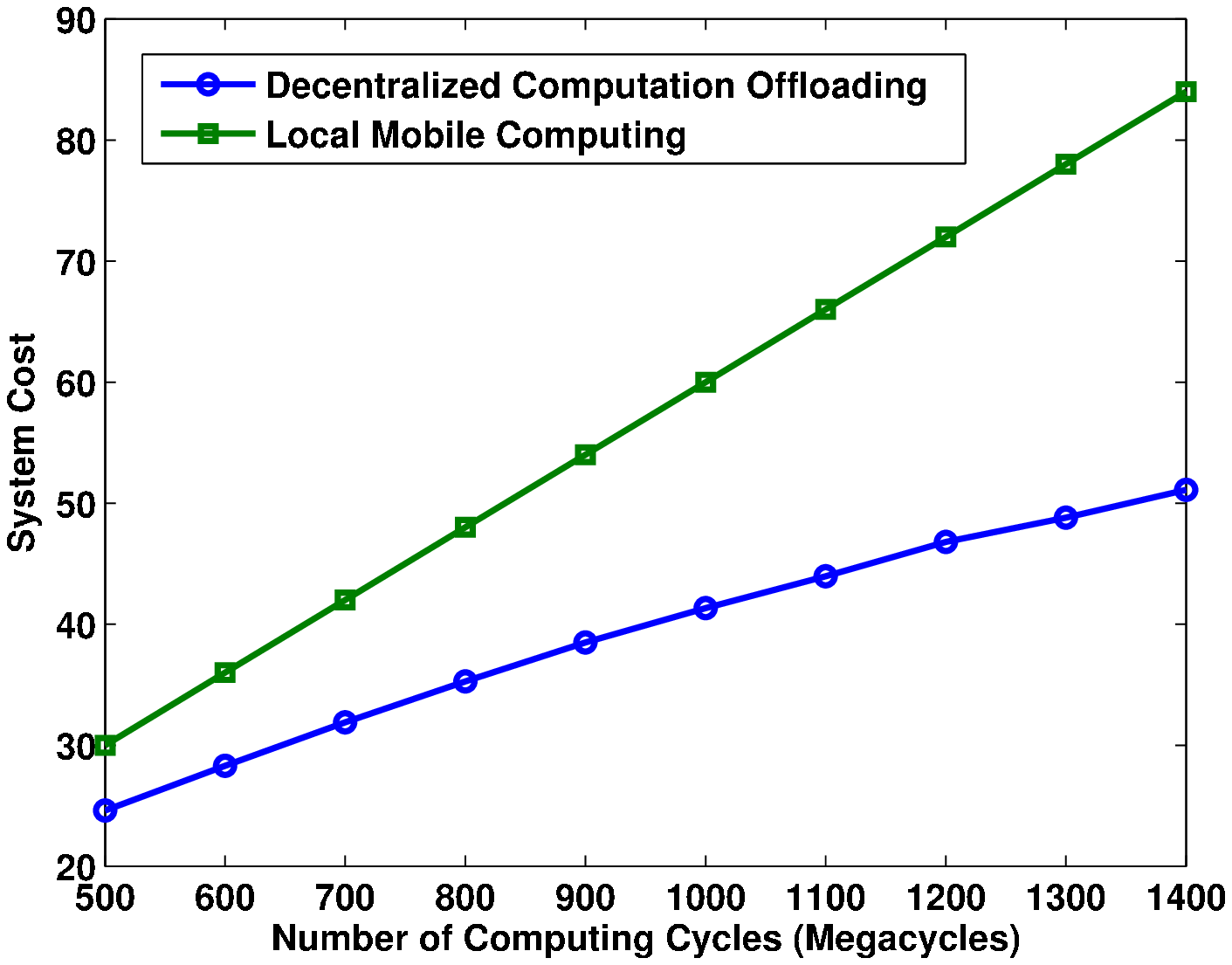}
\caption{\label{fig:System-wide-computing-cost}System-wide computing cost
with different number of CPU processing cycles }
\end{minipage}
\hfill
\begin{minipage}[t]{0.32\linewidth}
\centering
\includegraphics[scale=0.4]{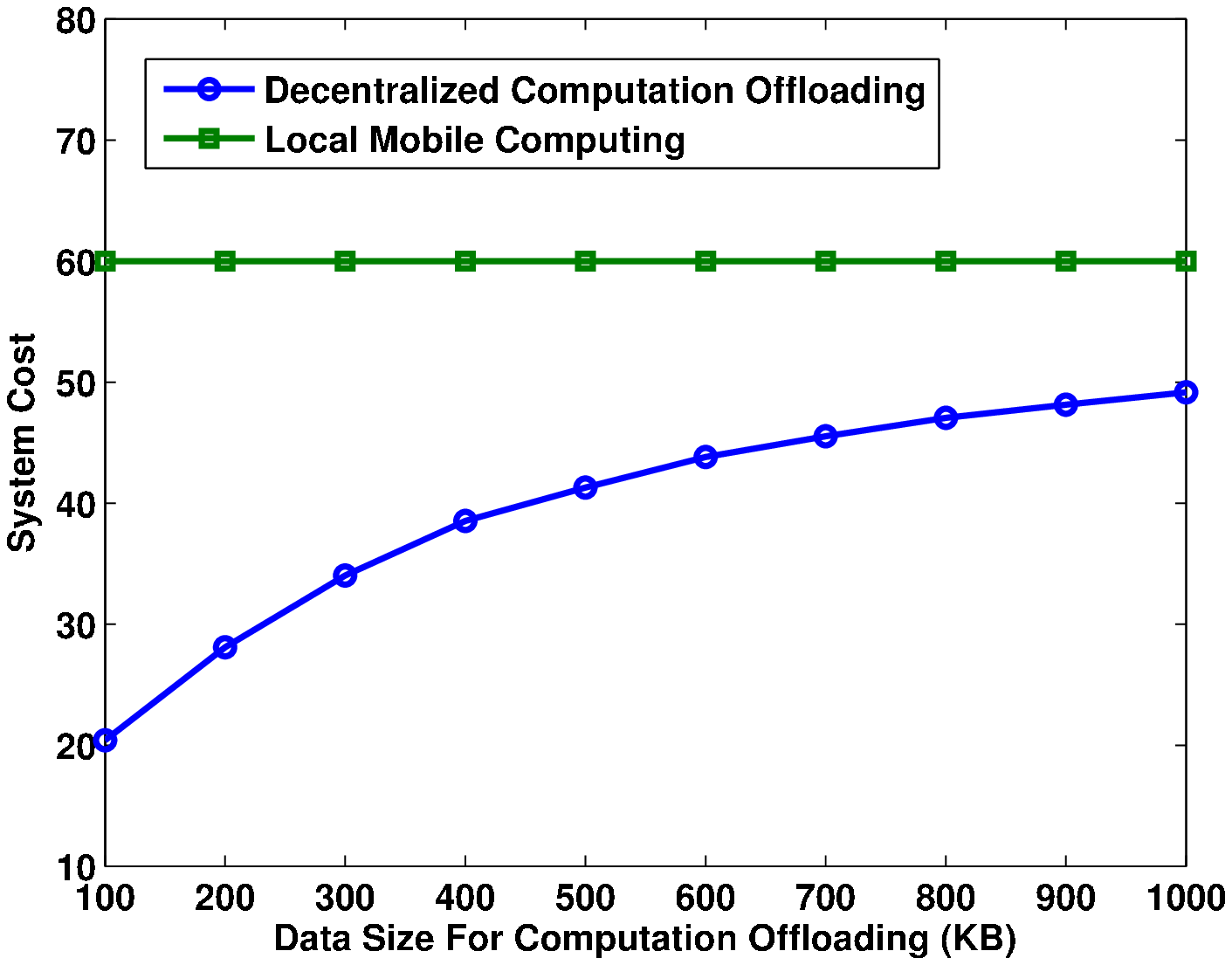}
\caption{\label{fig:System-wide-computing-cost-1}System-wide computing cost
with different data size for the computation offloading}
\end{minipage}
\end{figure*}

%
%
%\begin{figure}
%\centering
%\includegraphics[scale=0.5]{fig1}
%\caption{\label{fig:Dynamics-of-user}Dynamics of user cost by the decentralized
%computation offloading mechanism}
%\end{figure}
%
%
%\begin{figure}
%\centering
%\includegraphics[scale=0.5]{fig2}
%\caption{Dynamics of potential function \label{fig:Dynamics-of-potential}by
%the decentralized computation offloading mechanism}
%\end{figure}

\begin{figure*}[tt]
\begin{minipage}[t]{0.32\linewidth}
\centering
\includegraphics[scale=0.4]{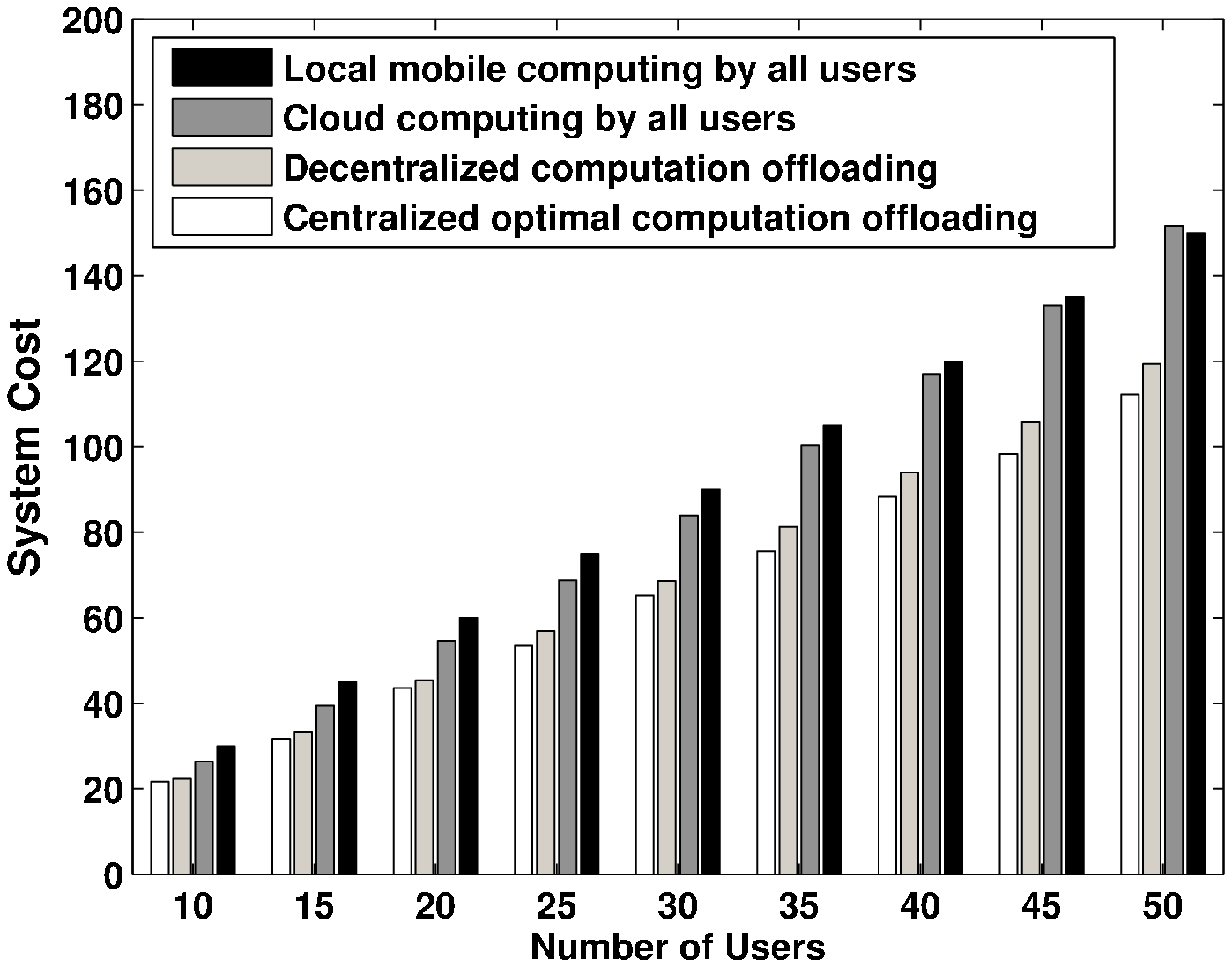}
\caption{\label{fig:Average-system-wide-computing}Average system-wide computing
cost}
\end{minipage}
\hfill
\begin{minipage}[t]{0.32\linewidth}
\centering
\includegraphics[scale=0.41]{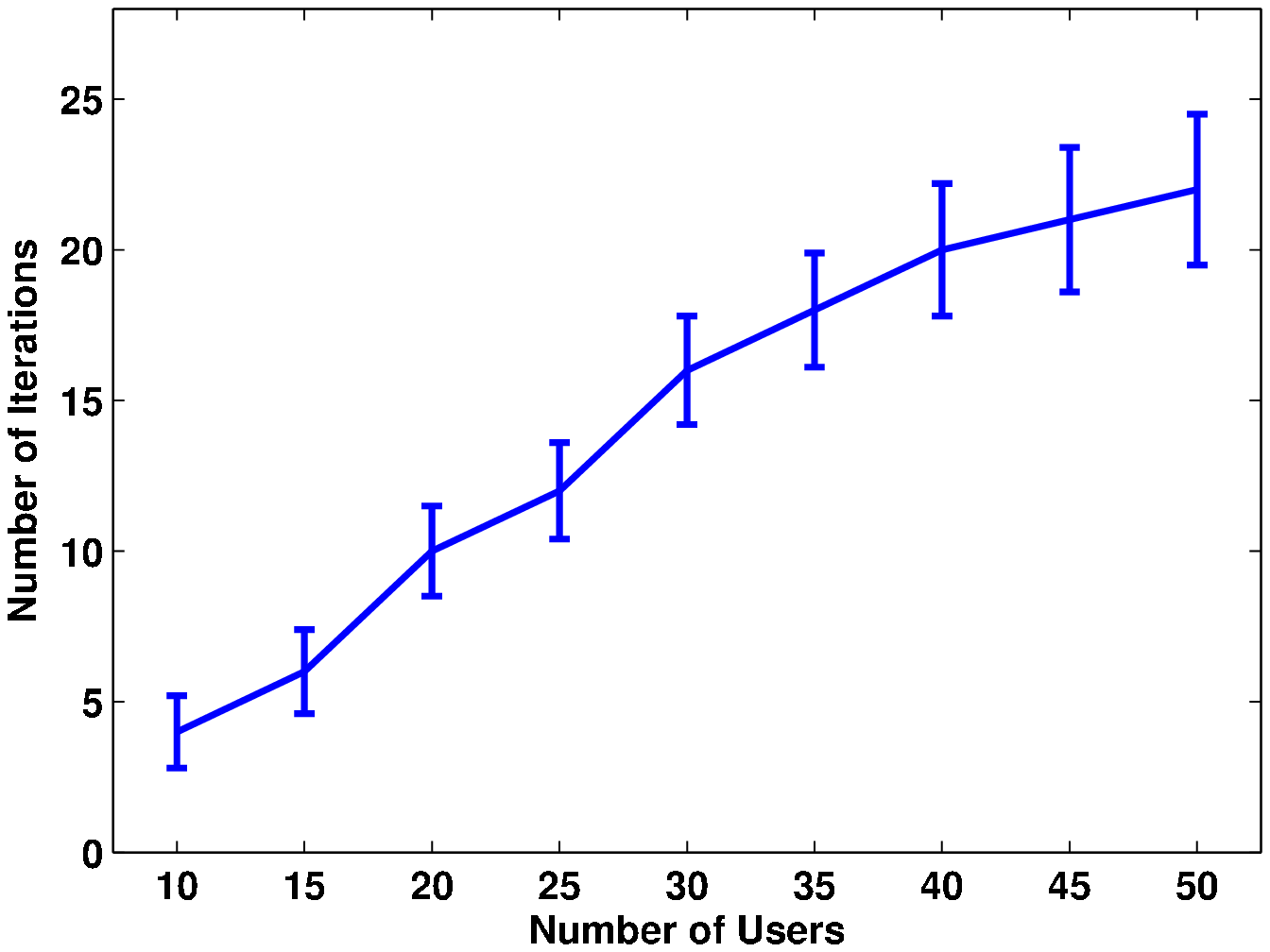}
\caption{\label{fig:Convergence-time-of}Number of iterations by decentralized
computation offloading mechanism}
\end{minipage}
\hfill
\begin{minipage}[t]{0.32\linewidth}
\centering
\includegraphics[scale=0.4]{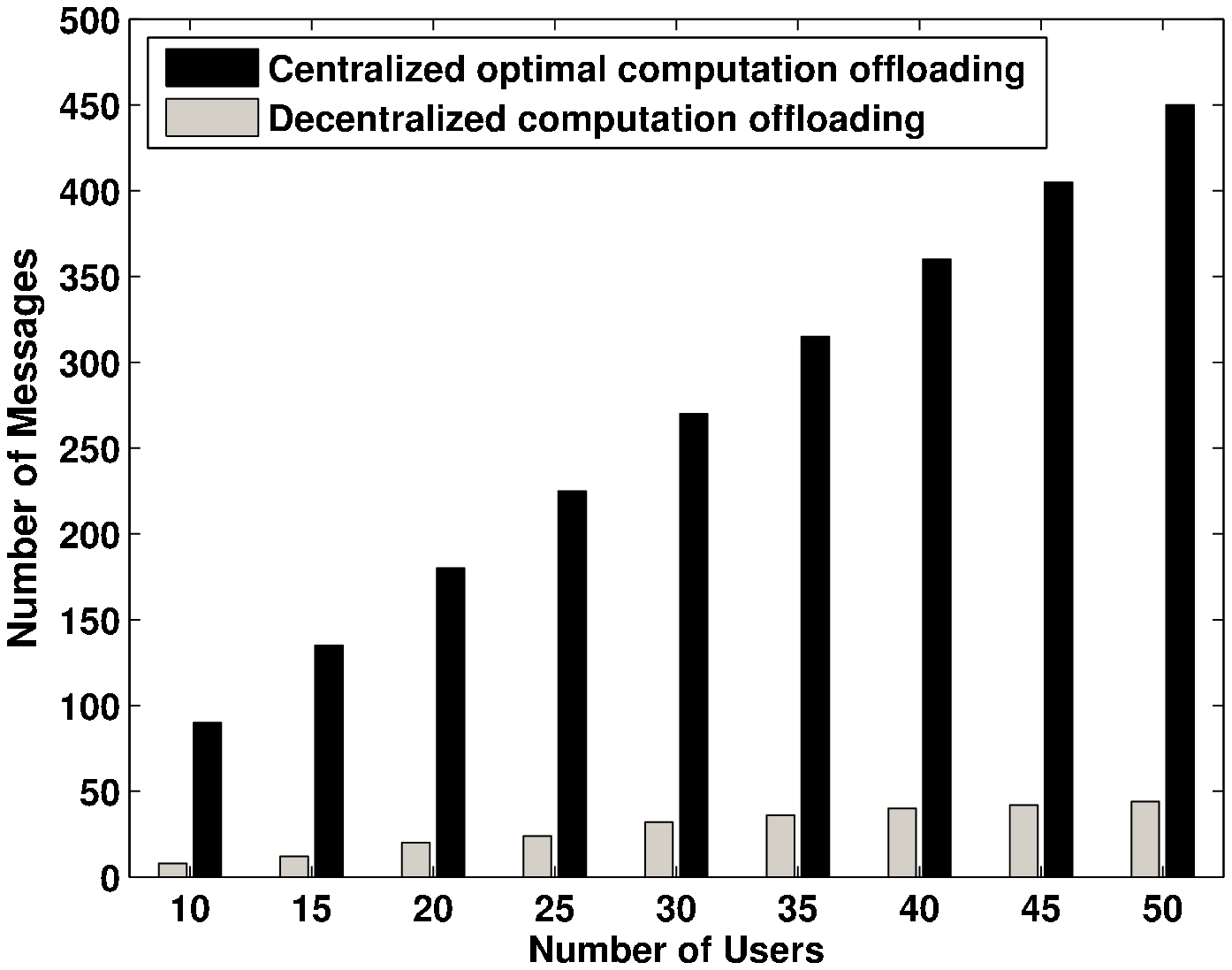}
\caption{\label{fig:Messages}Number of controlling and signaling messages by the centralized optimal and decentralized computation offloading mechanisms}
\end{minipage}
\end{figure*}

%\begin{figure}
%\centering
%\includegraphics[scale=0.5]{fig5}
%\caption{\label{fig:Average-system-wide-computing}Average system-wide computing
%cost}
%\end{figure}
%
%
%\begin{figure}
%\centering
%\includegraphics[scale=0.5]{fig6}
%\caption{\label{fig:Convergence-time-of}Number of iterations by decentralized
%computation offloading mechanism}
%\end{figure}
%
%
%\begin{figure}
%\centering
%\includegraphics[scale=0.5]{fig7}
%\caption{\label{fig:Messages}Number of controlling and signaling messages by the centralized optimal and decentralized computation offloading mechanisms}
%\end{figure}

We first show the dynamics of mobile device users' computation cost $V_{n}(\boldsymbol{a})$ by the
proposed decentralized computation offloading mechanism in Figure
\ref{fig:Dynamics-of-user}. We see that the mechanism can keep mobile
users' cost decreasing and converge to an equilibrium. To verify that
the convergent equilibrium is a Nash equilibrium, we further show
the dynamics of the potential function value $\Phi(\boldsymbol{a})$
of the decentralized computation offloading game in Figure \ref{fig:Dynamics-of-potential}.
It demonstrates that the proposed decentralized computation offloading
mechanism can lead the potential function of the game to the minimum
point, which is a Nash equilibrium according to the property of potential
game.

To investigate the impact of computation size on decentralized computation
offloading, we then implement the simulations with different number
of CPU processing cycles $D_{n}$ required for completing the computing
task. Upon comparison, we also implement the local mobile computing
solution such that all the mobile device users compute their tasks locally
on the mobile devices. The results are shown in Figure \ref{fig:System-wide-computing-cost}.
We see that the system-wide computing cost $\sum_{n\in\mathcal{N}}V_{n}(\boldsymbol{a})$
by decentralized computation offloading and local mobile computing
solutions increases as the number of CPU processing cycles $D_{n}$
increases. However, the system-wide computing cost $\sum_{n\in\mathcal{N}}V_{n}(\boldsymbol{a})$
by decentralized computation offloading increases much slower than
that of local mobile computing. This is because that as the number
of CPU processing cycles $D_{n}$ increases, more mobile device users choose
to utilize the cloud computing via computation offloading to mitigate
the heavy cost of local computing.

To evaluate the impact of communication data size on the decentralized
computation offloading, we next implement the simulations with different
data size for computation offloading $B_{n}$ in Figure \ref{fig:System-wide-computing-cost-1}.
We observe that the system-wide computing cost $\sum_{n\in\mathcal{N}}V_{n}(\boldsymbol{a})$
by decentralized computation offloading as the data size for computation
offloading $B_{n}$ increases, due to the fact that a larger data
size requires higher overhead for computation offloading via wireless
communication. Moreover, we see that the system-wide computing cost
$\sum_{n\in\mathcal{N}}V_{n}(\boldsymbol{a})$ by decentralized computation
offloading increases slowly when the data size for computation offloading
$B_{n}$ is large. This is because that the data size for computation
offloading $B_{n}$ is large, more mobile device users choose to compute
the tasks locally on the mobile devices, in order to avoid the heavy
cost of computation offloading via wireless access.

To benchmark the performance of the decentralized computation offloading
mechanism, we further implement the system-wide computing cost minimization
solution by centralized optimization, i.e., $\max_{\boldsymbol{a}}\sum_{n\in\mathcal{N}}V_{n}(\boldsymbol{a}).$
Notice that the centralized optimization solution requires the complete
information of all mobile device users, such as the details of computing
tasks, the transmission power, the channel gain, and the CPU frequency
of all mobile devices. While the decentralized computation offloading
mechanism only requires each mobile device user to measure its received interference
and make the decision locally. We run experiments with the number
of $N=10,15,...,50$ mobile device users being randomly scattered over the
square area, respectively. We repeat
each experiment $100$ times and show the average system-wide computing
cost in Figure \ref{fig:Average-system-wide-computing}. We
see that the system-wide computing cost by all the computation
offloading solutions increases as the number of mobile
device users N increases. The proposed incentive compatible
computation offloading solution can reduce up-to $33\%$ and
$38\%$ computing cost over the solutions of all the users choosing
the local computing and choosing the cloud computing,
respectively. Compared with the centralized optimization
solution, the performance loss of the decentralized computation offloading mechanism
is less than $10\%$ in all cases. This demonstrates the efficiency of the proposed decentralized computation
offloading mechanism. We next evaluate the convergence time of the
decentralized computation offloading mechanism. Figure \ref{fig:Convergence-time-of}
shows that the average convergence time increases linearly with the
number of mobile device users $N$. This shows that the decentralized computation
offloading mechanism scales well with the size of mobile device users. This
is critical since computing the centralized optimal computation offloading
solution involves solving the integer programming problem (i.e., the
decision variables $a_{n}\in\{0,1\}$) and the computational complexity grows exponentially as the
number of mobile device users $N$ increases.

To  evaluate the controlling and signaling overhead reduction by the decentralized computation
offloading mechanism, we further show the number of controlling and signaling  messages exchanged among the mobile users and between the users and the cloud in Figure \ref{fig:Messages}. It demonstrates that the decentralized computation
offloading mechanism can reduce the number of controlling and signaling messages by at least $89\%$ over the centralized optimal computation offloading scheme in all cases.  This is because that for the decentralized computation offloading mechanism, a mobile user would exchange messages (for interference measurement and decision update announcement) only when it updates its computation decision. While for the centralized optimal computation offloading scheme, each mobile user needs to report all its local parameters to the cloud, including the transmission power, the channel gain, the background interference power, the local computation capability, and many other parameters. Moreover, in some application scenarios, due to privacy concerns some mobile users may be sensitive to the revealing of their local parameters and hence do not have the incentive to participate in the centralized optimal computation offloading scheme. While the decentralized computation offloading mechanism does not have this issue since each mobile user can make the computation offloading decision locally without exposing its local parameters.

%\begin{figure}
%\centering
%\includegraphics[scale=0.5]{fig3}
%\caption{\label{fig:System-wide-computing-cost}System-wide computing cost
%with different number of CPU processing cycles }
%\end{figure}
%
%
%\begin{figure}
%\centering
%\includegraphics[scale=0.5]{fig4}
%\caption{\label{fig:System-wide-computing-cost-1}System-wide computing cost
%with different data size for the computation offloading}
%\end{figure}

%\begin{figure}
%\centering
%\includegraphics[scale=0.5]{fig5}
%\caption{\label{fig:Average-system-wide-computing}Average system-wide computing
%cost}
%\end{figure}
%
%
%\begin{figure}
%\centering
%\includegraphics[scale=0.5]{fig6}
%\caption{\label{fig:Convergence-time-of}Number of iterations by decentralized
%computation offloading mechanism}
%\end{figure}
%
%
%\begin{figure}
%\centering
%\includegraphics[scale=0.5]{fig7}
%\caption{\label{fig:Messages}Number of controlling and signaling messages by the centralized optimal and decentralized computation offloading mechanisms}
%\end{figure}

\section{Conclusion}\label{sec:Conclusion}
In this paper, we consider the computation offloading decision making problem among mobile device users for mobile cloud computing and propose as a decentralized computation offloading game formulation. We show that the game always admits a Nash equilibrium for both cases of homogenous and heterogenous wireless access. We also design a decentralized computation offloading mechanism that can achieve a Nash equilibrium of the game and further quantify its price of anarchy. Numerical results demonstrate that the proposed mechanism is efficient and scales well as the system size increases.

For the future work, we are going to consider the more general case that mobile users may depart and leave dynamically within a computation offloading period. In this case, the user mobility patterns might play an important role in the problem formulation. 

\bibliographystyle{ieeetran}
\bibliography{MobileCloud}

%\begin{IEEEbiography}[{\includegraphics[width=1.2in,height=1.25in,clip,keepaspectratio]{./XuChen}}]{Xu Chen}
%(S'10-M'12) received the B.S. degree in electronic engineering from the South China University of Technology (Guangzhou, Guangdong, China) in 2008, and the Ph.D. degree in information engineering from the Chinese University of Hong Kong (Hong Kong, China) in 2012. Dr. Chen is currently a postdoctoral research fellow in the
%School of Electrical, Computer and Energy Engineering, Arizona State University (Tempe, Arizona, USA). His general research interests include cognitive radio networks, wireless resource allocation, network economics, mobile social networks, and game theory. He is the recipient of the Honorable Mention Award (the first runner-up of the best paper award) in IEEE international conference on Intelligence and Security Informatics (ISI), 2010.
%\end{IEEEbiography}

\newpage
\section{Appendix}
\subsection{Proof of Lemma \ref{lem:Given-the-strategies}}\label{pf1}
According to (\ref{eq:l3}), (\ref{eq:c4}) and (\ref{eq:V1}), we
obtain that
\begin{align*}
V_{n}(\boldsymbol{a})  = & Z_{n}^{c}(\boldsymbol{a})a_{n}+Z_{n}^{l}(1-a_{n})\\
  = & \left(\gamma_{n}^{T}\left(T_{n,off}^{c}(\boldsymbol{a})+T_{n,exe}^{c}\right) +\gamma_{n}^{E}E_{n}^{c}(\boldsymbol{a})\right)a_{n} \\
  & +\left(\gamma_{n}^{T}T_{n}^{l}+\gamma_{n}^{E}E_{n}^{l}\right)(1-a_{n})\\
  = & \left(\frac{\left(\gamma_{n}^{T}+\gamma_{n}^{E}P_{n}\right)B_{n}}{R_{n}(\boldsymbol{a})}+\gamma_{n}^{T}T_{n,exe}^{c}\right)a_{n} \\
  & +\left(\gamma_{n}^{T}T_{n}^{l}+\gamma_{n}^{E}E_{n}^{l}\right)(1-a_{n}).
\end{align*}
When user $n$'s best response $a_{n}^{*}=1$, by Definition \ref{Given-the-strategies},
we have that
\[
V_{n}(1,a_{-n}^{*})\leq V_{n}(0,a_{-n}^{*}),
\]
which implies that
\[
\frac{\left(\gamma_{n}^{T}+\gamma_{n}^{E}P_{n}\right)B_{n}}{R_{n}(\boldsymbol{a})}+\gamma_{n}^{T}T_{n,exe}^{c}\leq\gamma_{n}^{T}T_{n}^{l}+\gamma_{n}^{E}E_{n}^{l}.
\]
That is,
\[
R_{n}(\boldsymbol{a})\geq\frac{\left(\gamma_{n}^{T}+\gamma_{n}^{E}P_{n}\right)B_{n}}{\gamma_{n}^{T}T_{n}^{l}+\gamma_{n}^{E}E_{n}^{l}-\gamma_{n}^{T}T_{n,exe}^{c}}.
\]
According to (\ref{eq:R1}), we then have that
\begin{align*}
& \sum_{m\in\mathcal{N}\backslash\{n\}:a_{m}=1}P_{m}H_{m,s} \\
\leq & L_{n}\triangleq\frac{P_{n}H_{n,s}}{2^{\frac{\left(\gamma_{n}^{T}+\gamma_{n}^{E}P_{n}\right)B_{n}}{W\left(\gamma_{n}^{T}T_{n}^{l}+\gamma_{n}^{E}E_{n}^{l}-\gamma_{n}^{T}T_{n,exe}^{c}\right)}}-1}-\omega_{n}.
\end{align*}
Similarly, we can analyze the case when user $n$'s best response
$a_{n}^{*}=0$. \qed

\subsection{Proof of Lemma \ref{lem_homo}}\label{pf2}
According to (\ref{eq:h1}), for a user $i\in\mathcal{S}$, its received
interference
\begin{eqnarray*}
\sum_{m\in\mathcal{N}\backslash\{n\}:a_{m}=1}P_{m}H_{m,s} & = & \left(|\mathcal{S}|-1\right)K\\
 & \leq & L_{i}.
\end{eqnarray*}
Then it follows from Lemma \ref{lem:Given-the-strategies} that playing
the strategy $a_{i}=1$ is a best response. Similarly, for a user
$j\in\mathcal{N}\backslash\mathcal{S}$, we can show that playing
the strategy $a_{j}=0$ is a best response. Since all the users play
the best response towards each other, the proposed strategy profile
is a Nash equilibrium. \qed

\subsection{Proof of Theorem \ref{thm:homo}}\label{pf3}
Given the set of ordered mobile device users with $\frac{L_{1}}{K}\geq\frac{L_{2}}{K}\geq...\geq\frac{L_{N}}{K}$,
if $\frac{L_{1}}{K}<0$, it is easy to check that the beneficial cloud
computing group $\mathcal{S}=\varnothing$. In this case, all users
$n\in\mathcal{N}$ playing the strategy $a_{n}=0$ is a best response
and hence a Nash equilibrium.

If $\frac{L_{1}}{K}\geq0,$ we can construct
a beneficial cloud computing group $\mathcal{S}\neq\varnothing$ by
Algorithm \ref{alg:Beneficial-cloud-computing}. At the first step,
we set $\mathcal{S}=\{1\}$. It is easy to check that $|\mathcal{S}|=1\leq\frac{L_{1}}{K}+1,$
which satisfies the condition in (\ref{eq:h1}). At the step $2\leq t\leq N$,
we define that $\tilde{\mathcal{S}}=\mathcal{S}\cup\{t\}$. If $|\tilde{\mathcal{S}}|=t\leq\frac{L_{t}}{K}+1$,
we then set that $\mathcal{S}=\tilde{\mathcal{S}}$ and continue to
the next step. If $|\tilde{\mathcal{S}}|=t>\frac{L_{t}}{K}+1$, we
can stop and obtain the beneficial cloud computing group as $\mathcal{S}=\{1,..,t-1\}$.
This is because that: 1) $\mathcal{S}$ satisfies that $|\mathcal{S}|\leq\frac{L_{t-1}}{K}+1$
(otherwise we cannot proceed to the step $t$), which implies that
the condition in (\ref{eq:h1}) is satisfied, i.e., $|\mathcal{S}|\leq\frac{L_{t-1}}{K}+1\leq...\leq\frac{L_{1}}{K}+1$;
2) since $|\tilde{\mathcal{S}}|=t>\frac{L_{t}}{K}+1$, we have $|\mathcal{S}|=t-1>\frac{L_{t}}{K}$,
which implies that the condition in (\ref{eq:h2}) is satisfied, i.e.,
$|\mathcal{S}|>\frac{L_{t}}{K}\geq...\geq\frac{L_{N}}{K}$. Note that
the total number of available steps of Algorithm \ref{alg:Beneficial-cloud-computing}
is bounded by $N$. A beneficial cloud computing group $\mathcal{S}\neq\varnothing$
hence must can be obtained. \qed

\subsection{Proof of Theorem \ref{thm:The-general-decentralized}}\label{pf4}
We first show that that $V_{k}(1,a_{-k})<V_{k}(0,a_{-k})$ implies
$\Phi(1,a_{-k})<\Phi(0,a_{-k})$ for a user $k$. For this case, according
to (\ref{eq:l3}), (\ref{eq:c4}) and (\ref{eq:V1}), the condition
$V_{k}(1,a_{-k})<V_{k}(0,a_{-k})$ implies that
\begin{equation}
\sum_{m\ne k}P_{m}H_{m,s}I_{\{a_{m}=1\}}<L_{k}.\label{eq:P5}
\end{equation}
Furthermore, according to (\ref{eq:p4}), we know that
\begin{eqnarray*}
 &  & \Phi(1,a_{-k})\\
 & = & \frac{1}{2}P_{k}H_{k,s}\sum_{m\ne k}P_{m}H_{m,s}I_{\{a_{m}=1\}}\\
 &  & +\frac{1}{2}\sum_{k\ne m}P_{m}H_{m,s}I_{\{a_{m}=1\}}P_{k}H_{k,s}\\
 &  & +\frac{1}{2}\sum_{n\ne k}\sum_{m\ne n,k}P_{n}H_{n,s}P_{m}H_{m,s}I_{\{a_{n}=1\}}I_{\{a_{m}=1\}}\\
 &  & +\sum_{n\ne k}P_{n}H_{n,s}L_{n}I_{\{a_{n}=0\}},
\end{eqnarray*}
and
\begin{eqnarray*}
 &  & \Phi(0,a_{-k})\\
 & = & \frac{1}{2}\sum_{n\ne k}\sum_{m\ne n,k}P_{n}H_{n,s}P_{m}H_{m,s}I_{\{a_{n}=1\}}I_{\{a_{m}=1\}}\\
 &  & +P_{k}H_{k,s}L_{k}+\sum_{n\ne k}P_{n}H_{n,s}L_{n}I_{\{a_{n}=0\}},
\end{eqnarray*}
which implies that
\begin{eqnarray}
 &  & \Phi(1,a_{-k})-\Phi(0,a_{-k})\nonumber \\
 & = & \frac{1}{2}P_{k}H_{k,s}\sum_{m\ne k}P_{m}H_{m,s}I_{\{a_{m}=1\}}\nonumber \\
 &  & +\frac{1}{2}\sum_{k\ne m}P_{m}H_{m,s}I_{\{a_{m}=1\}}P_{k}H_{k,s}\nonumber \\
 &  & -P_{k}H_{k,s}L_{n}\nonumber \\
 & = & P_{k}H_{k,s}\sum_{m\ne k}P_{m}H_{m,s}I_{\{a_{m}=1\}}\nonumber \\
 &  & -P_{k}H_{k,s}L_{k}.\label{eq:P6}
\end{eqnarray}
Combining (\ref{eq:P5}) and (\ref{eq:P6}), we have that
\[
\Phi(1,a_{-k})<\Phi(0,a_{-k}).
\]
 Similarly, for the case that $V_{k}(0,a_{-k})<V_{k}(1,a_{-k})$ for
a user $k$, we can also show that $\Phi(0,a_{-k})<\Phi(1,a_{-k})$. \qed

\subsection{Proof of Theorem \ref{thm:The-PoA-of}}\label{pf5}
Let $\hat{\boldsymbol{a}}\in\Upsilon$ be an arbitrary Nash equilibrium.
We must have that $V_{n}(\hat{\boldsymbol{a}})\leq Z_{n}^{l}$. Otherwise,
if $V_{n}(\hat{\boldsymbol{a}})>Z_{n}^{l}$, the mobile device user $n$
can always improve by choosing $a_{n}=0$ and experiencing a cost
of $Z_{n}^{l}$, which contradicts with the fact that $\hat{\boldsymbol{a}}$
is a Nash equilibrium. Thus, we have that $\sum_{n\in\mathcal{N}}V_{n}(\hat{\boldsymbol{a}}))\leq\sum_{n=1}^{N}Z_{n}^{l}$.

For an arbitrary computation offloading decision profile $\boldsymbol{a}=(a_{n},a_{-n})\in\prod_{n=1}^{N}\mathcal{A}_{n}$,
if $a_{n}=0$, we have $V_{n}(\boldsymbol{a})=Z_{n}^{l}$. If $a_{n}=1$,
we have that
\begin{eqnarray*}
R_{n}(\boldsymbol{a}) & = & W\log_{2}\left(1+\frac{P_{n}H_{n,s}}{\omega_{n}+\sum_{m\in\mathcal{N}\backslash\{n\}:a_{m}=1}P_{m}H_{m,s}}\right)\\
 &  & \leq W\log_{2}\left(1+\frac{P_{n}H_{n,s}}{\omega_{n}}\right),
\end{eqnarray*}
 which implies that
\begin{eqnarray*}
Z_{n}^{c}(\boldsymbol{a}) & = & \frac{\left(\gamma_{n}^{T}+\gamma_{n}^{E}P_{n}\right)B_{n}}{R_{n}(\boldsymbol{a})}+\gamma_{n}^{T}T_{n,exe}^{c}\\
 &  & \leq\frac{\left(\gamma_{n}^{T}+\gamma_{n}^{E}P_{n}\right)B_{n}}{W\log_{2}\left(1+\frac{P_{n}H_{n,s}}{\omega_{n}}\right)}+\gamma_{n}^{T}T_{n,exe}^{c}\\
 &  & =\overline{Z_{n}^{c}}.
\end{eqnarray*}
Thus, we know that $V_{n}(\boldsymbol{a})\ge\min\{Z_{n}^{l},\overline{Z_{n}^{c}}\}$
and $\sum_{n\in\mathcal{N}}V_{n}(\boldsymbol{a})\geq\sum_{n=1}^{N}\min\{Z_{n}^{l},\overline{Z_{n}^{c}}\}$.
Then it follows that \begin{align*}\mbox{PoA} & =  \frac{\max_{\boldsymbol{a}\in\Upsilon}\sum_{n\in\mathcal{N}}V_{n}(\boldsymbol{a})}{\min_{\boldsymbol{a}\in\prod_{n=1}^{N}\mathcal{A}_{n}}\sum_{n\in\mathcal{N}}V_{n}(\boldsymbol{a})} \\ & \leq\frac{\sum_{n=1}^{N}Z_{n}^{l}}{\sum_{n=1}^{N}\min\{Z_{n}^{l},\overline{Z_{n}^{c}}\}}.\end{align*} \qed 

\end{document}